# AN EFFICIENT AND SECURE ROUTING PROTOCOL FOR MOBILE AD-HOC NETWORKS

**N. Ch. Sriman Narayana Iyengar**
School of Computing Science and Engineering, VIT University, Vellore, India
nchsniyengar48@gmail.com

&

**Syed Mohammad Ansar   Sachin kumar,  Piyush Nagar,  Siddharth Sharma ,  Akshay Atrey**
School of Information Technology and Engineering, VIT University, Vellore, India

*ABSTRACT*

*Efficiency and simplicity of random algorithms have made them a lucrative alternative for solving complex problems in the domain of communication networks. This paper presents a random algorithm for handling the routing problem in Mobile Ad hoc Networks [MANETS]. The performance of most existing routing protocols for MANETS degrades in terms of packet delay and congestion caused as the number of mobile nodes increases beyond a certain level or their speed passes a certain level. As the network becomes more and more dynamic, congestion in network increases due to control packets generated by the routing protocols in the process of route discovery and route maintenance. Most of this congestion is due to flooding mechanism used in protocols like AODV and DSDV for the purpose of route discovery and route maintenance or for route discovery as in the case of DSR protocol. This paper introduces the concept of random routing algorithm that neither maintains a routing table nor floods the entire network as done by various known protocols thereby reducing the load on network in terms of number of control packets in a highly dynamic scenario. This paper   calculates the expected run time of the designed random algorithm.*

*KEY WORDS*

*Random Algorithms, MANETS*

## 1.INTRODUCTION

MANETS [1, 2] are self organizing groups of mobile nodes which also act as router, connected by wireless links. MANETS do not have a centralized infrastructure (i.e.) there is no fixed central node to coordinate the task of routing.  Due to fixed wireless range the source node will use multiple hops to communicate to destination node.

With the recent technological advances in wireless communication and the increasing popularity of portable computing devices, wireless and mobile ad hoc networks are expected to play increasingly important role in future civilian and military settings where wireless access to wired backbone is either ineffective or impossible. Mobile ad hoc networks (MANETs) are composed of a set of stations (nodes) communicating through wireless channels, without any fixed backbone support. Applications of MANETs include, but are not limited to, military operations, security, emergency, and rescue operations, among other applications where intense utilization of a communication networks is available for a very limited time. However, frequent topology changes caused by node mobility make routing in wireless ad hoc networks a challenging problem. In addition, limited capabilities of mobile stations require a control on node congestion due to message forwarding and limited battery consumption. Mobility of mobile hosts introduces also new challenging problems that were not encountered in the design and implementation of conventional wireless and wired networks. A critical and challenging problem of mobile ad hoc networking and computing is how to fully cope with the special characteristics of the wireless and mobile ad hoc environment, make balanced use of computation and communication resources, and support user mobility. Most of the available literature in this emerging technology concentrates on physical and networking aspects of the subject. However, in most of the these studies, they have neglected the description of the fundamental design of distributed algorithms and have not discussed how to apply them to wireless and mobile ad hoc network environments. An important requirement for successful deployment of wireless and mobile ad hoc network-based applications is the careful evaluation of performance and investigation of alternative algorithms, prior to their implementation.

## 2. MAJOR ISSUES WITH EXISTING ROUTING PROTOCOLS FOR MANET'S

There are two types of routing protocols, proactive protocols and Reactive protocols. Proactive protocols periodically send control packets in the network to update routing tables. But as the network becomes more and more dynamic these control packets increase network congestion. Reactive protocols send control packets for route discovery upon demand from the sender. The performance of both proactive protocols as well as reactive protocols degrade when network becomes highly dynamic due to movement of nodes. This results in increment of packet delay and network congestion. Four typical routing protocols of ad hoc networks, which include one distance vector routing protocol DSDV[3] and three on-demand routing protocols AODV[4], DSR[5] and TORA[6]. DSDV is a table-driven protocol based on the classical Bellman-Ford mechanism. The improvements made to Bellman-Ford algorithm include freedom from loops in the routing table. Every mobile node in the network maintains a routing table in which all of the possible destinations within the network and the number of hops to each destination are recorded. While AODV, DSR and TORA share the on-demand behavior in that they initiate routing activity only in the presence of data packets in need of a route, many of their routing mechanism are different. AODV uses a table-driven routing framework and destination sequence numbers, DSR uses a source routing, whereas TORA uses a link reversal routing mechanism. Commonly, the latter three have a less routing load and the former has a less end-to-end delay. [7] and [8] have discussed and analyzed the performance of various existing routing protocols under different scenarios using simulation on the basis of following factors.

**Mobility** If the nodes starts to move with high speed, the performance of all known routing protocols get degraded. As the nodes will have to change their routing table (is they use proactive protocols) or it will become extremely difficult to deliver packet to the node because of the uncertainty of its position.

**Congestion**: Proactive protocols cannot be used, as the nodes keep on changing and so do the topology of the network, so the control messages (which will be generated continuously to keep track of the position of the nodes) will bring the network to its knees as a result of congestion.

**Power**: The mobile nodes run on battery power, so we have limited amount of power to route packets and deal with the changing topology of the network.

**Packet delivery time:** If we use reactive protocols the packet delivery time will increase since we have to find the route on demand. So the whole process will be too slow.

**Security:** Since the nodes are mobile and there is no central authority to authenticate nodes, node authentication is an important problem in ad hoc networks. Various kind of attacks are possible like packet sniffing, man in the middle attack, node impersonation etc.

**Number of nodes:** As the number of nodes increases its becomes difficult to route the packets, since its becomes difficult to keep track of the nodes.

**Packet Injection Rate:** It refers to the rate at which each node is injecting packet in the network. If this is quite high then the network becomes congested easily.

## 3. PROPOSED METHOD FOR ENHANCEMENT OF PERFORMANCE IN AD-HOC NETWORKS

A random walk, sometimes denoted RW, is a mathematical formalization of a trajectory that consists of taking successive random steps. The results of random walk analysis have been applied to computer science, physics, ecology, economics and a number of other fields as a fundamental model for random processes in time. For example, the path traced by a molecule as it travels in a liquid or a gas, the search path of a foraging animal, the price of a fluctuating stock and the financial status of a gambler can all be modeled as random walks. Specific cases or limits of random walks include the drunkard's walk and Lévy flight. Random walks are related to the diffusion models and are a fundamental topic in discussions of Markov processes. Several properties of random walks, including dispersal distributions, first-passage times and encounter rates, have been extensively studied.

Various different types of random walks are of interest. Often, random walks are assumed to be Markov, but other, more complicated walks are also of interest. Some random walks are on graphs, others on the line, in the plane, or in higher dimensions, while some random walks are on groups. Random walks also vary with regard to the time parameter. Often, the walk is indexed by the natural numbers, as in. However, some walks take their steps at random times, and in that case the position Xt is defined for.

In computing, a Las Vegas algorithm is a randomized algorithm that always gives correct results; that is, it always produces the correct result or it informs about the failure [10]. In other words, a Las Vegas algorithm does not gamble with the verity of the result; it only gambles with the resources used for the computation. A simple example is randomized quick sort, where the pivot is chosen randomly, but the result is always sorted. The usual definition of a Las Vegas algorithm includes the restriction that the expected run time always is finite, when the expectation is carried out over the space of random information, or entropy, used in the algorithm. Let A be a randomized algorithm that allows the answer "?". We say that A is a Las Vegas algorithm for a function F if for every input x,

1. $\Pr ob(A(x) = F(x)) >= \frac{1}{2}$ and

2. $\Pr ob(A(x) = "?") = 1 - \Pr ob(A(x) = F(x)) <= \frac{1}{2}$

The proposed protocol upon being called by the routing agent calls its receive function as a handler of the packet received from upper layer. The protocol uses special headers to find the immediately next hops by broadcasting specially designed small control packets with range determined as single hop. The benefit of using single hop is to reduce congestion in the network. As soon as the send function is called, a timer is started in the background to maintain the queue. This queue is used to store addresses of the single hop neighbors that reply to the broadcast. The addresses are enqueued until the timer expires. Once the timer expires check queue function is called, it deques the queue and check if the destination node is in the range. If the destination is in single hop range then the packet is delivered to it otherwise a built in random function is called to select a node at random from the queue and the packet is forwarded to it.

## 4. EXPECTED RUNNING TIME CALCULATION

Let,

$$X_{ij} = \text{Indicator random Variable} = \begin{cases} 1 & \text{if node } n_i \text{ and } n_j \text{ are neighbours} \\ 0 & \text{otherwise} \end{cases}$$

At time $T = 0$ nodes broadcast at one hop range to calculate the distance their neighbors.

$D_{ij}$ is the initial distance between node i and node j. Let there be a function f which gives the probability that distance between any two nodes $n_i$ and $n_j$ at time $T = t$ is $X = d$ is

$$f(X = d) = (\Pr ob(f(n_i, n_j, D_{ij}, T = t, X = d)))$$

At any time $T = t$ the probability that node $n_i$ is in range of $n_j$ is

$$F(X <= r) = \int_0^r (f(n_i, n_j, D_{ij}, T = t, X = d))$$

At any time $T = t$ the expected number of neighbors of any node $n_i$ is

$$E(n) = \sum_{j \in N-i} \int_0^r (f(n_i, n_j, D_{ij}, T = t, X = d)) * X_{ij}$$

Probability of selecting an edge $= \dfrac{1}{E(n)}$

Expected number of edges traversed in random walk

$$\sum \Pr obability * edjes \ traversed = F(X <= r) + \frac{1}{E(n)} * F(X <= 2r) * 2$$

$$+ \left[\frac{1}{E(n)}\right]^2 * F(X <= 3r) * 3 + \ldots \left[\frac{1}{E(n)}\right]^{k-1} * F(X <= kr) * k + \ldots$$

The function F follows poisson distribution because every case is independent and has a fixed and equal probability of happening. As the time increases the number of movements tend to be infinite thus satisfying the criteria of poisson distribution. [9] shows that if we iteratively use the random algorithm the resulting implementation will have a run time of $O(R_A)$ where, $R_A$ is the run time of any Las Vegas random algorithm.

## 5. SIMULATION AND ANALYSIS

To simulate any networking scenario in network simulator (NS-2) first of all the nodes are created via Tcl script and their initial positions are fixed. The traffic type to be simulated on the network is attached to the node via transport layer agent. On top of this transport layer agent the application layer agents like CBR or FTP are attached. Since the proposed protocol deals with only routing, the major concern is only with layer three simulations and implementation. With routing strategies (i.e.) the routing of the packets received from the upper layer or previous hop and delivers them at the destination node with no guarantee. Functionalities of other layers are taken care of by the upper and lower layer agents of NS-2. A node can only act as a source or sink at application layer level. The agent acting as source will generate traffic at application layer. This traffic is transferred to the transport layer which further attaches its own transport header to it. This packet is then received by RTagent after it has been de-multiplexed by the Demux agent. The RTagent is responsible for delivering it to the routing layer. The network layer agent, if busy enques this packet upon receiving whereas it calls the recv(receive) function if free. The receive function has two arguments one is the packet pointer and the other is the packet handler. The network layer calls the predetermined routing protocol that is registered using Tcl scenario.

## 6. RESULTS

Figure 1 shows output result on Xgraph which plots the packets delivered and packets lost vs time. This graph shows the performance of AODV protocol in the above generated scenario.

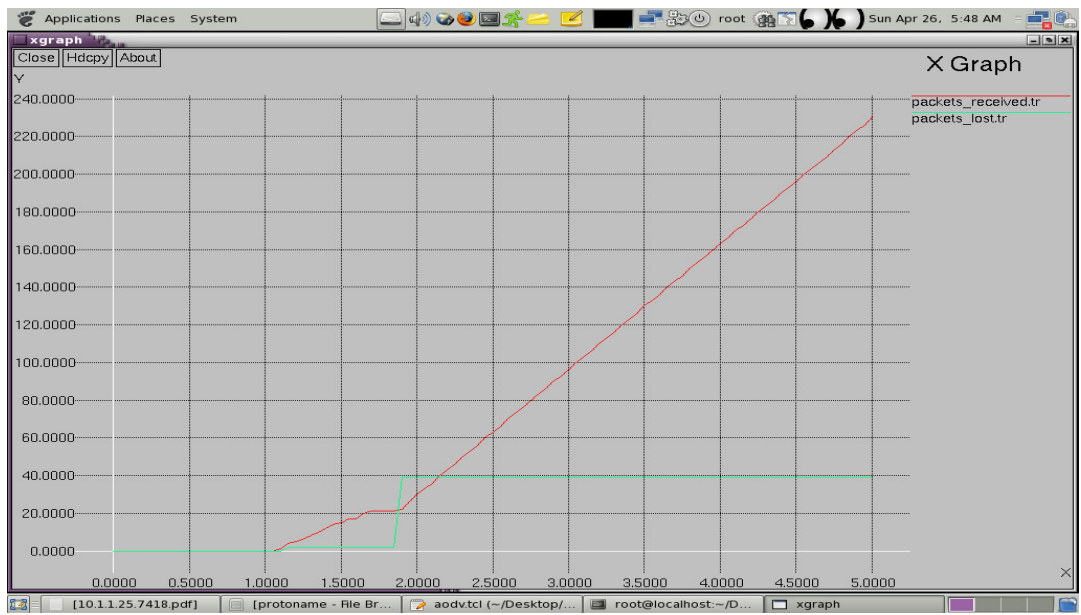

**Figure 1 Performance of AODV**

X axis represents time in seconds and y axis represents number of packets. The red line shows packets received and blue line depicts number of packet dropped. The proposed protocols graph in Figure 2 shows that when the scenario is dynamic then the performance of protocol is relatively far better than AODV. AODV is an on-demand protocol or reactive protocol. Whenever a node wants to send a packet the destination address is checked in the routing table and packet is forwarded appropriately. Whenever destination node moves or some intermediate node moves the table becomes stale and the network is flooded with control packets to construct a new routing table. When nodes are frequently mobile the performance of AODV degenerates in terms of packet delay and congestion. The network is simulated with 6 wireless nodes in which the destination node is not in direct range of source. At time t=1 node 0 which is the source node starts sending CBR traffic to node 5 which is the destination node. At the same time node 5 starts to move from coordinates (600,200) to (100,200). It takes nearly 1 sec to reach its destination. When the performance of AODV and random walk is analyzed in this scenario in the time interval 1 sec to 2 sec (i.e.) when the destination node is mobile, it is observed that random walk outperforms AODV in terms of number of packets received. Since a routing table is not maintained, the delay is quite less to forward the packet randomly when compared to AODV which spends a lot of time in route discovery and route maintenance via flooding the network with control packets. The graph between number of packets versus time, which is produced at the end of the simulation infers that in time interval seconds the packets dropped by AODV rises exponentially whereas the increase in slope of packets dropped in random walk is far more less than that of AODV.

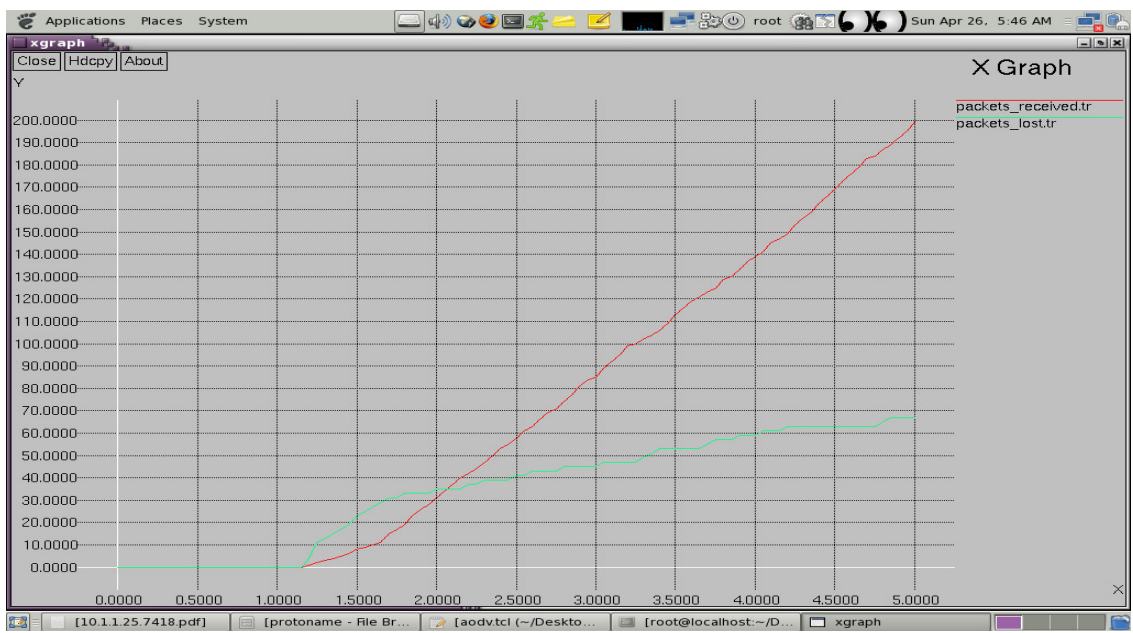

**Figure 2 Performance of Proposed Protocol**

## 7. CONCLUSION

In this paper is proposed an iterative random algorithm to forward packets in a Mobile Ad hoc Network [MANET]. The expected run time of the algorithm is calculated. The protocol implementation on MANET scenario using ns2 simulator is simulated and the results are compared with that of AODV. AODV is regarded as the best protocol for highly dynamic MANETs. The proposed protocol outperforms AODV in terms number of packets received and dropped when the network was dynamic, i.e. when the nodes were mobile. The results are compared using the graphs where the packets dropped and packet received v/s time for both protocols are plotted. The analysis of the graphs shows that the proposed protocol performs better when the nodes are mobile.

During the interval of (1sec, 2sec) the packets dropped for AODV shows a sudden rise. AODV's graph for packets dropped is nearly straight line in this interval representing a sudden rise in the number of packets dropped in a small interval. When the nodes are moved with higher speed the nature of the graph remains unchanged stating that whenever the nodes move, the proposed protocol outperforms AODV.

The simulations were conducted using limited computational facilities; therefore simulations were performed only on limited number of nodes. The nature of random routing protocol was studied and experimented using small dynamic network which had its limitations in the form of number and types of traffic packets, number of nodes and speed of the nodes. Another limitation being the results were obtained using simulator which does not necessarily represents networks in real time implementations. The limited computing resources of a single machine simulating the whole network scenario can give results deviating from that of real time scenarios.

The proposed protocol lays the foundation for further work on applications of iterative random routing protocols in highly dynamic MANET scenarios. The immediate future work can include implementation of this protocol on real time ad-hoc networks. The proposed protocol finds application in Vehicular Ad-hoc Networks [VANETs] where nodes (Vehicles on road) are highly mobile. VANETs are finding application in automated traffic management and vehicular navigation in modern vehicles where vehicles can communicate with each other without any network infrastructure using ad-hoc protocols. The proposed protocol will outperform AODV, which is currently regarded as the best protocol in such scenarios. Our proposed protocol can also find application in military communications in war scenario where security of the nodes cannot be compromised. Our protocol will perform better as the nodes are highly dynamic and will be secure for communication as there is no single point of attack in routing as nodes are selected randomly. The proposed protocol is secure from man-in-the-middle attack because of oblivious nature of packet forwarding. Since no routing table is maintained there is no fixed path between source and destination, therefore there is no single point of attack for adversary. *Oblivious* routing is a type of distributed routing suitable for dynamic packet arrivals. In oblivious routing, the path for a newly injected packet is selected in a way that it is not affected by the path choices of the other packets in the network. [6] Gives an existential result that shows that for any network there exists an oblivious routing algorithm with congestion within factor log3 *n* from that of the optimal off-line centralized algorithm, where *n* is the number of nodes. This oblivious algorithm constructs a path by choosing a logarithmic number of random intermediate nodes in the network.